\documentclass[11pt]{article}
\usepackage{amssymb}

\newcommand{\BE}{\begin{equation}}
\newcommand{\EE}{\end{equation}}
\newcommand{\BA}{\begin{eqnarray}}
\newcommand{\EA}{\end{eqnarray}}
\newcommand{\vol}{{\sf V}}

\begin{document}
\begin{titlepage}

\vspace*{1mm}
\begin{center}

            {\LARGE{\bf Modern Michelson-Morley experiments \\
and gravitationally-induced anisotropy of $c$ }}

\vspace*{14mm}
{\Large  M. Consoli }
\vspace*{4mm}\\
{\large
Istituto Nazionale di Fisica Nucleare, Sezione di Catania \\
c/o Dipartimento di Fisica dell' Universit\`a \\
Via Santa Sofia 64, 95123 Catania, Italy}
\end{center}
\begin{center}
{\bf Abstract}
\end{center}
The recent, precise Michelson-Morley experiment performed by
 M\"uller et al. suggests 
a tiny anisotropy of the speed of light.
I propose a quantitative explanation of the observed effect based on the 
interpretation of gravity as a density fluctuation of the Higgs condensate.
\vskip 35 pt
\end{titlepage}


{\bf 1.}~The aim of this Letter is to propose a quantitative explanation 
of the tiny anisotropy of the speed of light suggested by the recent, 
precise Michelson-Morley experiment of M\"uller et al.
\cite{muller}.
 Their result can be conveniently expressed in the form 
\BE 
\label{B}
                          B^{\rm exp}= (-3.1 \pm 1.6)\cdot 10^{-9}
\EE 
where $B$ enters the Robertson-Mansouri-Sexl 
\cite{mansouri} parametrization for the two-way speed of light
($c=2.9979...10^{10}$ cm/sec) 
\BE
\label{ctheta}
          {{  \bar{c} }\over{c}}= 1 - (A +B \sin^2 \theta) 
{{v^2}\over{c^2}}
\EE
in a reference frame $S'$ that 
moves with speed $v$ with respect to a preferred frame $\Sigma$, 
$\theta$ denoting the angle between the direction of $v$ and the direction 
of the light beam.
\vskip 10 pt
{\bf 2.}~In order to explain the experimental result reported in
Eq.(\ref{B}), as a first step, I'll adopt
the tentative idea that light propagates in 
 a medium with refractive index ${\cal N}_{\rm medium} > 1$  so
that there is a small Fresnel's drag coefficient
$1-{{1}\over { {\cal N}^2_{\rm medium}}}\ll 1$. This provides
a general framework 
to analyze any Michelson-Morley type of experiment 
(see refs.\cite{cahill,pagano}). In our case, where the medium is the 
vacuum itself, the physical interpretation of ${\cal N}_{\rm vacuum}$
will represent a second step and provide a quantitative estimate 
to be compared with Eq.(\ref{B}). 

In this
perspective, I'll start introducing 
an isotropical speed of light 
\BE
u= {{c}\over{\cal{N}_{\rm medium} }}
\EE
that refers to the ideal case of a medium that extends to infinity in all
spatial directions. Real experiments, however, are performed in finite
portions of medium that might even reduce to
just fill the arms of an interferometer. In this situation, an observer 
placed on the earth's surface has no argument to think that light should
propagate with the same velocity $u$ in all directions. 

However, one may adopt the point of view that any observed anisotropy is
due to the earth's motion with respect to a preferred frame $\Sigma$ 
where light
propagates isotropically. In this case, if $\Sigma$ were identified with 
the cosmic background radiation, one expects 
the relevant value of the earth's 
velocity to be $v_{\rm earth}\sim 365$ km/sec \cite{cobe}. 
By adopting this point of view, and recalling that
Lorentz transformations are valid both in Special and Lorentzian Relativity, 
for a frame $S'$, moving with respect to $\Sigma$ 
with velocity ${\bf{v}}$, light will be seen to propagate at a speed 
($\gamma= 1/\sqrt{ 1- {{ {\bf{v}}^2}\over{c^2}} }$) 
\BE
\label{uprime}
  {\bf{u}}'= {{  {\bf{u}} - \gamma {\bf{v}} + {\bf{v}}
(\gamma -1) {{ {\bf{v}}\cdot {\bf{u}} }\over{v^2}} }\over{ 
\gamma (1- {{ {\bf{v}}\cdot {\bf{u}} }\over{c^2}} ) }}
\EE
To second order in $v/u$, one obtains
($\theta$ denotes the angle between ${\bf{v}}$ and ${\bf{u}}$)
\BE
  {{u'(\theta)}\over{u}}= 1- \alpha {{v}\over{u}} -\beta {{v^2}\over{u^2}}
\EE
where
\BE
   \alpha = (1-  {{1}\over{ {\cal N}^2_{\rm medium} }} ) \cos \theta + 
{\cal O} ( ({\cal N}^2_{\rm medium}-1)^2 )
\EE
\BE
\beta = (1- {{1}\over{ {\cal N}^2_{\rm medium} }} ) P_2(\cos \theta) +
{\cal O} ( ({\cal N}^2_{\rm medium}-1)^2 )
\EE
with $P_2(\cos \theta) = {{1}\over{2}} (3 \cos^2\theta -1)$.

Thus, the two-way speed of light is 
\BE
\label{twoway}
{{\bar{u}'(\theta)}\over{u}}= {{1}\over{u}}~ {{ 2  u'(\theta) u'(\pi - \theta) }\over{ 
u'(\theta) + u'(\pi - \theta) }}= 1- {{v^2}\over{c^2}} ( A + B \sin^2\theta) 
\EE
where 
\BE 
   A= {\cal N}^2_{\rm medium} -1 + {\cal O} ( ({\cal N}^2_{\rm medium}-1)^2 )
\EE
and 
\BE
\label{BTH}
     B= -{{3}\over{2}} 
({\cal N}^2_{\rm medium} -1 )
+ {\cal O} ( ({\cal N}^2_{\rm medium}-1)^2 )
\EE
In this way, using the experimental values
${\cal N}_{\rm air}\sim 1.00029$ or
${\cal N}_{\rm helium}\sim 1.000036$, one can re-analyze 
\cite{cahill,pagano} the classical `ether-drift' experiments. 
For instance, by defining
\BE
v_{\rm earth} \sqrt{ {\cal N}^2_{\rm medium} -1}= v_{\rm obs}
\EE
(and an in-air-operating optical system)
one predicts $v_{\rm obs} \sim 9 $ km/sec for
$v_{\rm earth}\sim 365 $ km/sec, in good agreement with
Miller's results \cite{miller}.

\vskip 10 pt
{\bf 3.}~To compare with Eq.(\ref{B}) I'll now
try to provide a quantitative estimate of ${\cal N}_{\rm vacuum}$, to be used
in Eq.(\ref{BTH}), starting from
the idea of a `condensed' vacuum, as generally accepted
 in modern elementary particle
physics. Indeed, in the physically relevant case of the Standard Model,
the situation can be summarized saying \cite{thooft} that 
 "What we experience as 
empty space is nothing but the configuration of the Higgs field that has the
lowest possible energy. If we move from field jargon to particle jargon, this 
means that empty space is actually filled with Higgs particles. They have 
Bose condensed." 

In this case, where the
condensing quanta are just neutral spinless particles 
(the `phions' \cite{mech}), 
the translation from `field jargon to particle jargon', 
amounts to establish a well defined functional relation (see ref.\cite{mech})
$n=n(\phi^2)$ between the average particle density 
$n$ in the ${\bf{k}}=0$ mode and the average value of the scalar 
field $\langle \Phi \rangle=\phi$. 
Thus, Bose condensation is just a consequence of
the minimization condition of the 
 effective potential $V_{\rm eff}(\phi)$. This has absolute
minima at some values $\phi =\pm v \neq 0$ for which $n(v^2)=\bar{n}\neq 0$
\cite{mech}. 

The symmetric phase, where $\phi=0$ and $n=0$, will eventually be 
re-established at a phase
transition temperature $T=T_c$. This, in the Standard Model, is so high 
that one can safely approximate the ordinary vacuum as a zero-temperature 
system. Thus, the vacuum might be compared to 
a quantum Bose liquid, a medium
where bodies can flow without any apparent friction, as in superfluid $^4$He,
in agreement with the experimental results.

On the other hand, the condensed particle-physics vacuum, while certainly different
from the ether of classical physics, is also different from
the `empty' space-time of Special Relativity which is assumed at the base of 
axiomatic quantum field theory.  
Therefore, following this line of thought, 
 the macroscopic occupation of the same quantum
state (${\bf{k}}=0$ in a given reference frame)
 can represent the operative construction
of a `quantum ether'
  whose existence might be detected through a
precise `ether-drift' experiment.

On a more formal ground we observe that the coexistence of 
{\it exact} Lorentz covariance and vacuum condensation in {\it effective}
quantum field theories is not so trivial. In fact, as a consequence of the
violations of locality at the energy scale fixed by the
ultraviolet cutoff $\Lambda$ \cite{salehi}, 
one may be faced with non-Lorentz-covariant 
{\it infrared} effects that depend on the vacuum structure.

To indicate this type of infrared-ultraviolet connection, originating 
from vacuum condensation in effective quantum field theories, 
Volovik \cite{volo1} has introduced a very appropriate name: 
reentrant violations of special relativity in the low-energy corner. 
In the simplest case of spontaneous symmetry breaking in a 
$\lambda\Phi^4$ theory, where the condensing
quanta are just neutral spinless particles, the `reentrant' effects reduce
to a small shell of three-momenta, say $|{\bf{k}}| < \delta$, where the 
energy spectrum deviates from a Lorentz-covariant form. Namely, 
by denoting $M_H$ as the typical energy scale associated with the 
Lorentz-covariant part of the energy spectrum, one finds
${{\delta}\over{M_H}} \to 0$ only when 
 ${{M_H}\over{\Lambda}}\to 0$. 

The basic ingredient to detect such `reentrant' 
effects in the broken phase consists in a purely
quantum-field-theoretical result:
the connected zero-four-momentum propagator $G^{-1}(k=0)$ 
is a two-valued function
\cite{legendre,pmu}. In fact, besides the well known solution
$G^{-1}_a(k=0)= M^2_H$, one also finds
$G^{-1}_b(k=0)= 0$. 

The b-type of solution
corresponds to processes where assorbing (or emitting)
a very small 3-momentum ${\bf{k}} \to 0$
does not cost any finite energy. This situation is well
known in a condensed medium, where a small 
3-momentum can be coherently distributed 
among a large number of elementary constituents, and corresponds to 
the hydrodynamical regime of density fluctuations whose
wavelengths $2\pi/|{\bf{k}}|$ are {\it larger} 
than $r_{\rm mfp}$, 
the mean free path for the elementary constituents. 

This interpretation \cite{weak,hierarchy}
 of the gap-less branch, which is very natural on the base of general 
arguments, is unavoidable in a superfluid medium. In fact, 
 "Any quantum liquid consisting of particles with integral 
spin (such as the liquid isotope $^4$He) must certainly have a spectrum of
this type...In a quantum Bose liquid, elementary excitations with small 
momenta ${\bf{k}}$  (wavelengths large compared with distances between atoms) 
correspond to ordinary hydrodynamic sound waves, i.e. are phonons. This 
means that the energy of such quasi-particles is a linear function of their
momentum" \cite{pita}. In this sense, a superfluid 
vacuum provides for ${\bf{k}} \to 0$ a universal picture.
This result does not depend
on the details of the short-distance interaction and even on the nature
of the elementary constituents. 
For instance, the same coarse-grained description is found in 
superfluid fermionic vacua \cite{volo2} that, as compared to the Higgs 
vacuum, bear the same relation of superfluid $^3$He to superfluid $^4$He.

Thus there are
two possible types of excitations with the same quantum numbers but
different energies when the 3-momentum ${\bf{k}} \to 0$: 
a single-particle
massive one, with ${E}_a({\bf{k}}) \to M_H$, and a collective
gap-less one with 
${E}_b({\bf{k}}) \to 0$. In this sense, the situation is very similar to
superfluid $^4$He, 
where the observed energy spectrum is due to the peculiar
transition from the `phonon branch' to the `roton branch' at a momentum scale 
$|{\bf{k}}_o|$ where
${E}_{\rm phonon}({\bf{k}}_o) \sim 
{E}_{\rm roton}({\bf{k}}_o)$. 
The analog for the scalar condensate amounts to
an energy spectrum with the following limiting behaviours :

~~~i) ${E}({\bf{k}}) \to {E}_b({\bf{k}}) \sim c_s |{\bf{k}}|$   
       ~~~~~~~~~~~~~~~~~~~~ for ${\bf{k}}\to 0 $

~~~ii) ${E}({\bf{k}}) \to {E}_a({\bf{k}}) \sim M_H+ {{ {\bf{k}}^2 }\over{2 M_H}}$
       ~~~~~~~~~~ for $|{\bf{k}}| \gtrsim \delta $

where the 
characteristic momentum scale $\delta \ll M_H$, at which
$E_a(\delta)\sim E_b(\delta)$, marks the transition from collective to 
single-particle excitations. This occurs for
\BE
\delta \sim 1/r_{\rm mfp} 
\EE
 where \cite{kine,seminar}
\BE
r_{\rm mfp} \sim {{1}\over{ \bar{n} a^2}}
\EE
is the phion mean free path, 
for a given value of the phion density $n=\bar{n}$ and a
given value of the phion-phion scattering length $a$. In terms of the 
same quantities, one also finds \cite{mech}
\BE
               M^2_H \sim \bar{n} a
\EE
giving the trend of the dimensionless ratios ($\Lambda \sim 1/a$)
\BE
\label{golden}
{{\delta}\over{M_H}} \sim {{M_H}\over{\Lambda}} \sim \sqrt{ \bar{n}a^3} \to 0
\EE
in the continuum limit where $a \to 0$ and the mass scale $\bar{n}a$ is
held fixed. 

By taking into account the above results, the physical
decomposition of the scalar field in 
the broken phase can be conveniently
expressed as (phys=`physical') \cite{physical}
\BE
\label{phime2}
\Phi_{\rm phys}(x) = v_R + {h}(x) + {H}(x)
\EE 
with 
\BE
\label{hh}
{h}(x)=
\sum_ { | {\bf {k}}| < \delta }  
\frac{1} { \sqrt{2 \vol {E}_k } } 
\left[  \tilde{h}_{\bf k}    {\rm e}^ { i ({\bf k}.{\bf x} -{E}_k t) } + 
(\tilde{h}_{\bf k})^{\dagger} {\rm e}^{-i ({\bf k}.{\bf x} -{E}_k t)} 
\right]
\EE
and
\BE
\label{H2}
{H}(x)=
\sum_{ |{\bf {k}}| > \delta }  
\frac{1} { \sqrt{2 \vol {E}_k } } 
\left[  \tilde{H}_{\bf k}    {\rm e}^ { i ({\bf k}.{\bf x} -{E}_k t) } + 
(\tilde{H}_{\bf k})^{\dagger} {\rm e}^{-i ({\bf k}.{\bf x} -{E}_k t)} 
\right]
\EE
where $\vol$ is the quantization volume and 
${E}_k=c_s|{\bf{k}}|$ for $|{\bf{k}}| < \delta$ while
${E}_k=\sqrt{{\bf{k}}^2 + M^2_H}$ for $|{\bf{k}}| > \delta$. Also, 
$c_s \delta \sim M_H$. 

Eqs.(\ref{phime2})-(\ref{H2}) replace
the more conventional relations
\BE
\label{conve1}
\Phi_{\rm phys}(x)= v_R + H(x)
\EE 
where
\BE
\label{conve2}
{H}(x)=
\sum_{ {\bf {k}} }
\frac{1} { \sqrt{2 \vol {E}_k } } 
\left[  \tilde{H}_{\bf k}    {\rm e}^ { i ({\bf k}.{\bf x} -{E}_k t) } + 
(\tilde{H}_{\bf k})^{\dagger} {\rm e}^{-i ({\bf k}.{\bf x} -{E}_k t)} 
\right]
\EE
with ${E}_k=\sqrt{{\bf{k}}^2 + M^2_H}$. 
Eqs.(\ref{conve1}) and (\ref{conve2})
are reobtained in the limit 
${{\delta}\over{M_H}} \sim {{M_H}\over{\Lambda}} \to 0$ 
where the wavelengths associated to
$h(x)$ become infinitely large in units of the physical scale set by 
$\xi_H=1/M_H$.  In this limit, where for any finite value of
${\bf{k}}$ the broken phase has
only massive excitations, one recovers an exactly Lorentz-covariant theory.
\vskip 10 pt
{\bf 4.}~In conclusion, for finite values of $\Lambda$ there are long-wavelength density 
fluctuations of the vacuum and
Lorentz-covariance is not exact. Therefore, in the presence of such 
effects, one can try to 
detect the existence of the scalar condensate through a precise 
`ether-drift' experiment.  
To this end, I observe that
 a simple physical interpretation of the long-wavelength density 
fluctuation field 
\BE
\varphi(x)\equiv {{h(x)}\over{v_R}}
\EE
 has been proposed in 
refs.\cite{weak,hierarchy}. Introducing $G_F\equiv 1/v^2_R$ 
and choosing
the momentum scale $\delta$ as 
\BE
\label{choice}
                      \delta= \sqrt{  {{G_N M^2_H }\over{G_F}}  }
\EE
($G_N$ being the Newton constant) one obtains the identification
\BE
\varphi(x)=U_N(x) + {\rm const.}
\EE
$U_N(x)$ being the Newton potential. Indeed, with the choice in 
Eq.(\ref{choice}), to first order in $\varphi$ and 
in the limits of slow motions, the equations of
motion for $\varphi$ reduce to 
the Poisson equation for the Newton potential $U_N$ \cite{weak,hierarchy}
so that the deviations from
Lorentz covariance are of gravitational strength. If, as in the Standard Model,
$G_F$ is taken to be the Fermi constant one then
finds $\delta\sim 10^{-5}$ eV and $r_{\rm mfp}\sim 1/\delta=
{\cal O}(1)$ cm. As anticipated, the variation of
$\varphi(x)$ takes place over distances 
that are larger than $r_{\rm mfp}$ and thus
infinitely large on the elementary particle scale. Also, by introducing 
$M_{\rm Planck}= {{1}\over{ \sqrt {G_N} }}$, and using Eqs.(\ref{golden}) and
(\ref{choice}), one finds $\Lambda = q_H M_{\rm Planck}$ with 
$ q_H= \sqrt{ G_F M^2_H}= {\cal O}(1)$, or 
$a \sim 1/\Lambda \sim 10^{-33}$ cm.

At the same time, to first order, the observable 
effects of $\varphi$ can be re-absorbed \cite{hierarchy}
into an effective metric structure
\BE
ds^2= (1+ 2\varphi) dt^2 - (1-2\varphi)(dx^2 +dy^2 +dz^2)
\EE
that agrees with the first approximation to the line element of
 General Relativity \cite{rosen,weinberg}. 
In this perspective, 
the space-time curvature arises from two sources: i) 
a re-scaling of the length and time units associated with the modification
of any particle mass 
and ii)  a refractive index for
light propagation 
\BE
\label{nphi}
            {\cal N}_{\rm vacuum}\sim  1- 2\varphi
\EE
needed to preserve the basic particle-wave duality which is intrinsic 
in the nature of light. 

Now, for a centrally symmetric field, and up to a constant, 
one has 
           $ \varphi(R) =- {{G_N M}\over{c^2 R}}$.
Therefore, 
$\varphi_{\rm earth}\sim -0.7\cdot 10^{-9}$ (for $M=M_{\rm earth}$ and 
$R=R_{\rm earth}$) so that, using Eq.(\ref{BTH}), I would estimate
\BE
\label{theor}
                 B^{\rm th} \sim -4.2 \cdot 10^{-9}
\EE
in good agreement with the experimental result in 
Eq.(\ref{B}).
\vskip 10 pt
{\bf 5.}~Summarizing:  the vacuum is not `empty' so that
one should check the consistency between 
exact Lorentz
covariance and vacuum condensation in effective
quantum field theories. For the specific case of the scalar condensate,
 the non-locality
associated with the presence of the ultraviolet cutoff will also
show up at long wavelengths in the form of non-Lorentz-covariant 
density fluctuations 
associated with a scalar function $\varphi(x)$. 

If, on the base of refs.\cite{weak,hierarchy}, these long-wavelength
effects are naturally interpreted in terms of the Newton potential $U_N$
(with the identification $\varphi=U_N+ {\rm const.}$), one obtains
the weak-field space-time curvature of General Relativity
and a refractive index ${\cal N}_{\rm vacuum}\sim 1-2\varphi$. This value of 
${\cal N}_{\rm vacuum}$, leading to the prediction in Eq.(\ref{theor}),
can help to understand the experimental result Eq.(\ref{B})
obtained by M\"uller et al. \cite{muller}.


\begin{thebibliography} {99}
\bibitem{physical}
By `physical' we mean that the normalization of the vacuum field
$v_R$ is such that 
the quadratic shape of $V_{\rm eff}(\phi_R)$ at $\phi_R=v_R$ is precisely 
given by the physical Higgs mass $M^2_H$.
\end{thebibliography}
\end{document}